# Planar Waveguide Circulating Gaussian Beam Resonators on a Silicon Photonic Chip

SIEGFRIED JANZ*, SHURUI WANG, RUBIN MA, JEAN LAPOINTE, AND MARTIN VACHON

*Quantum and Nanotechnology Research Centre, National research Council Canada, Building M-50, 1200 Montreal Road, Ottawa, Ontario, Canada, K1A 0R6*
*siegfried.janz@nrc-cnrc.gc.ca*

**Abstract:** A Si slab waveguide resonator design with a circulating Gaussian-like cavity mode is described and characterized, for both all-pass and add-drop configurations and several different input/output coupling strengths. The circulating beam propagates in a slab waveguide with no lateral confinement. Three straight mirrors and one curved mirror define a folded two-dimensional Gaussian cavity. Light is coupled to and from the resonator by beam splitters formed by a narrow gap between a cavity mirror and the input slab waveguides. The coupling is determined by the gap width and is wavelength independent and lossless. For a L=100 μm path length cavity, resonance line widths of $\lambda_{3dB}$ = 5 pm with Q-values of Q = 310000 were measured. The resonator drop spectrum exhibited a comb of almost identical resonance lines across a 100 nm tuning range. This resonator design is capable of broadband operation and is less susceptible to sidewall roughness and defect scattering induced loss and mode splitting, when compared with Si rings formed from narrow single mode waveguides.

## 1. Introduction

Waveguide ring resonators are a well-established optical element in silicon photonic integrated optics [1,2]. Ring resonators provide a comb-like transmission spectrum similar to that of a Fabry-Perot etalon in bulk-optics. They have been employed as filters for blocking or transmitting specific wavelengths of light in optical communications [3,4], in low power optical-modulators [5,6], and biological, chemical and temperature sensors [7-11].

A conventional Si ring resonator consists of single mode waveguide configured in a closed loop optical path. Light is coupled between the ring and the single mode input and output waveguides using 2×2 waveguide directional couplers. The common Si waveguide thickness is 220 nm, with a width no more than 500 nm to ensure single mode operation. The waveguides are patterned into the top Si layer of a silicon-on-insulator wafer, and the buried oxide (i.e., $SiO_2$) layer forms the lower cladding. The upper cladding can be $SiO_2$, air, or any other sufficiently low index material as required for the application.

In practice, fabricating Si ring resonators with a desired specification and performance level can be challenging because of the small waveguide dimensions and high refractive index of the Si waveguide core. The etched sidewall roughness and defects on the scale of a few nm cause waveguide loss which limits the achievable Q-factor, while back-scattering into counterpropagating modes produces resonance line shape distortion and line splitting [12,13]. The directional coupler usually has a narrow wavelength envelope with a 3dB bandwidth of 30 nm to 50 nm, depending on details of the design [14]. Small variations in the directional coupler gap cause the coupling efficiency peak to deviate from the target wavelength. This can be problematic when the ring is combined with other wavelength dependent circuit elements such as surface grating couplers. When the passband envelopes of different elements do not overlap in wavelength, the entire chip can be rendered unusable. While the high local intensity in a single mode Si waveguide can lead to interesting nonlinear effects, these same effects are a problem in applications where precise resonance wavelength reproducibility is essential. For example, power dependent optical self-heating can be a significant source of reading error in Si ring photonic thermometers [15].

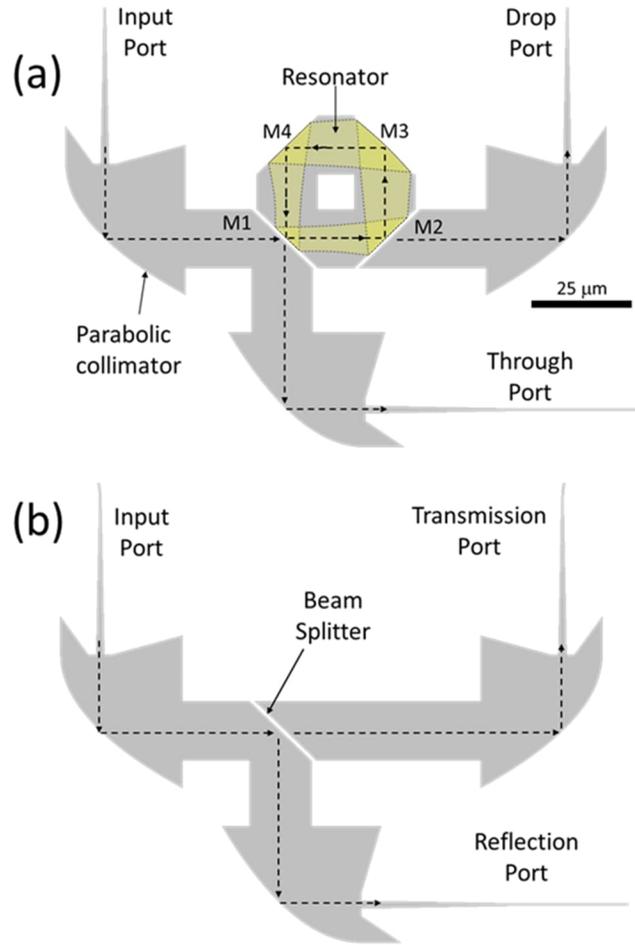

Fig. 1. (a) Layout of the add-drop CGMR resonator. The Si slab waveguide areas are shaded in grey. The dashed arrows indicate the light propagation direction, and the yellow shaded area indicates the approximate shape of the circulating Gaussian mode in the cavity. Mirror M3 is curved to achieve Gaussian beam confinement. The identical all-pass resonator is obtained by removing the Si waveguide area and collimator to the right of mirror M2. (b) The test structure for independently measuring reflection and transmission of the beam splitters.

In this work, we attempt to mitigate these problems by using circulating Gaussian beam resonator (CGBR) design that does not use narrow Si waveguides and directional couplers. The resonator is formed in a Si slab waveguide with no lateral confinement. The light is confined to the resonator using shaped mirrors which operate in the total internal reflection (TIR) regime, and hence are essentially lossless. Rather than directional couplers, the add and drop ports of the resonator are simple beam splitters based on evanescent coupling across a narrow gap.

Resonator confinement is achieved by curving one of the mirrors to form a planar Gaussian mode that circulates in the cavity. Propagation loss due to sidewall scattering is largely eliminated, and input and output coupling is almost wavelength independent across the C-band. The optical mode is several micrometers wide, so local self-heating is much less than in conventional single mode Si waveguides. These devices should be particularly suitable for sensor applications.

## 2. CGMR Design

Figure 1(a) shows the layout of the prototype CGBR resonator characterized in this work. The resonator is formed by an arrangement of 15 μm wide slab waveguides fabricated on an SOI wafer with a 220 nm thick Si waveguide layer on a 2 μm thick buried oxide layer. The upper waveguide cladding is air. Four mirrors are fabricated by etching a vertical wall through the Si waveguide layer. These mirrors are positioned at each of the four corners of a 25 μm × 25 μm square that defines a L= 100 μm long circulating light path. Light is coupled with the resonator through a beam splitter created by a narrow gap between the corner mirror M1 and the input/output slab waveguide. This beam splitter forms the input and through port, in analogy with the through port of a conventional ring resonator [1,2]. A small fraction of the incident light is coupled into the resonator by evanescent coupling across the gap, while the remainder is reflected and re-focused to the through port output waveguide. To create a resonator in an add-drop configuration, a second identical beam splitter is placed at mirror M2 to couple light to the drop port. In the all-pass configuration, the beam splitter and output waveguide beyond M2 are removed. All-pass and add-drop devices were fabricated with four different beam splitter gap distances ranging from D=200 nm to D=600 nm. Given the calculated Si slab waveguide effective index of $N_{eff}$ = 2.74, the 45° incident angle of light at each mirror or beam splitter is well beyond the TIR angle of $\theta_{TIR}$ = 21° at the Si waveguide /air interface, so the mirrors and beam splitters are in principle lossless. Since TIR is a result of overall wave vector conservation between the slab waveguide and air/cladding, the vertical sidewall angle and profile should not change the overall splitter and mirror loss. However, roughness and defects along the side wall may still result in scattering loss. For each beam splitter gap, a diagnostic structure shown in Fig. 1(b) consisting of a beam splitter with only transmission and reflection output ports is used to independently measure the splitter transmission, reflection, and loss.

The light circulating within the resonator only interacts with etched Si side walls at the mirrors and splitters, and is otherwise not laterally confined. The circulating light would thus normally diffract with propagation distance, resulting in cavity loss and poor resonator performance. To create a confined resonator mode, mirror M3 is curved with a R = 115 μm radius of curvature. At an incident angle of 45°, this corresponds to an effective focal length $F_e$ = R/√2 of approximately 81 μm. This curved mirror shapes the beam into circulating resonator mode that resembles a planar 100 μm long Gaussian cavity mode folded into a square optical path with a beam waist at mirror M1. Applying the Gaussian beam formulae [16], the predicted $1/e^2$ half-width at M1 is $\omega_0$ = 3.6 μm, corresponding to full width of 7.2 μm. The maximum beam width at mirror M3 is 8.6 μm. The radius of curvature of M3 was chosen so that the resonator beam width at M1 is comparable to the incident beam width projected on M1 by the input collimator. The slab waveguide width of the resonator is 15 μm, so the circulating beam has negligible overlap with the etched sidewalls except at the corner mirrors and beam splitters.

In order to couple light into the resonator, the incident slab mode arriving at mirror M1 should be collimated and ideally have a field lateral profile that matches the resonator beam waist shape and size at M1. Any mismatch will result in a coupling loss. A parabolic mirror beam collimator or beam expander etched into a Si slab waveguide can be used the expand the beam from a single mode photonic wire waveguide to a collimated beam of any desired width [17,18]. Since the parabolic collimator also operates in the TIR regime, this beam transformation is theoretically lossless. In our device, we use a F=10 μm focal length parabola to create a collimated beam with diameter of approximately 8 µm. The incident light in launched into the parabola from a 500 nm wide single mode Si waveguide input, which is tapered out to a 2 µm aperture at the parabola focus. An identical collimator focuses the light exiting the resonator at mirror M1 into the single mode through port output waveguide. For the add-drop resonator, the light coupled out of the resonator at the M2 beam splitter is focused into the drop port waveguide by a third collimator.

If the field envelope matching between incident and output slab beams and the internal resonator modes at mirrors M1 and M2 is perfect, the couplers will function identically to a 2×2 directional coupler [1,2]. The directional coupler through and cross coupling coefficients are simply replaced by the beam splitter reflectivity and transmission. In this prototype CGMR device, the launched waveguide mode profile at the input of the collimator section is only qualitatively similar to a Gaussian mode shape, and the parabolic collimator itself introduces a small asymmetry in the collimated beam profile. As a result, the standard ring resonator equations [1, 2] will not apply without modification to include mismatch losses at input/through port mirror M1. Although there will also be a mode overlap mismatch at the drop port beam splitter at M2, the drop signal itself is linearly proportional to the power circulating inside the resonator, and can therefore still be used to characterize the internal resonator performance.

## 3. Experiment

In the experiments, TE polarized light is coupled to the Si chip from a single mode fiber with a facet polished at an 8° angle, using surface grating couplers [19,20]. An identical arrangement is used to collect the output light. The measured peak coupling efficiency of the fabricated grating couplers was -4 dB at λ=1520 nm, with a 3 dB wavelength band pass of approximately 50 nm. The grating couplers are designed to minimize back reflections, but a high frequency ripple noise of up to ±1dB in the measured device spectra indicates that residual back-reflections from the gratings or other sources remain present in the optical path. A 500 nm wide single mode Si waveguide connects the grating couplers to the input and output ports of the resonator shown in Fig. 1. The waveguide loss of these connecting waveguides was -1.8 dB/cm, as determined by comparing the insertion loss of three waveguides of lengths 6.1 mm, 17 mm and 31 mm.

All the data was obtained using a wavelength tunable laser with a tuning range from λ = 1460 nm to 1580 nm. The incident in-fiber power was 100 µW. The measured spectra presented in this work have all been normalized by the transmission spectrum of a 500 nm wide straight reference waveguide of the same length as the input and output waveguides of the resonators, and terminated by the same input and output grating couplers. The results therefore represent the insertion loss of the resonators and test structures alone. The insertion loss uncertainty is estimated to be ±1 dB due to small variations in waveguide loss and the grating coupler efficiencies, pass band shapes and peak wavelengths at different locations due to cross-chip fabrication variations.

Four identical versions of the resonator in Fig. 1 were characterized, with beam splitter gap widths of 200 nm, 300 nm, 400nm and 500 nm, for structures K, L, M, and N respectively as indicated in Table 1. Both an all-pass and add-drop version of each resonator were fabricated. For each splitter gap, a corresponding test structure as in Fig. 1(b) was used to determine the reflectivity T and transmission R of each beam splitter. T and R are calculated as the ratios T=$P_1$/($P_1$+$P_2$) and R=$P_2$/($P_1$+$P_2$), where $P_1$ and $P_2$ are the powers in output transmission port 1 and the reflection port 2. The transmission spectrum of each splitter is shown in Fig. 2. The reflection port spectrum is not shown, since it was indistinguishable from the reference waveguide spectrum within the measurement uncertainty. The combined insertion loss of the two parabolic collimators and beam splitter in the R port path are therefore better than -1 dB. Both R and T are approximately constant across the measured wavelength range. As noted previously, the small wavelength variations of the transmission spectra in Fig. 2 are likely an artifact of small differences in the grating couplers for the R and T ports. The transmission values at $\lambda = 1520$ nm are given in Table 1. They range from $T = 6.6 \times 10^{-3}$ for a 200 nm gap (Device K) to $T = 7.9 \times 10^{-5}$ for a 500 nm gap (Device N).

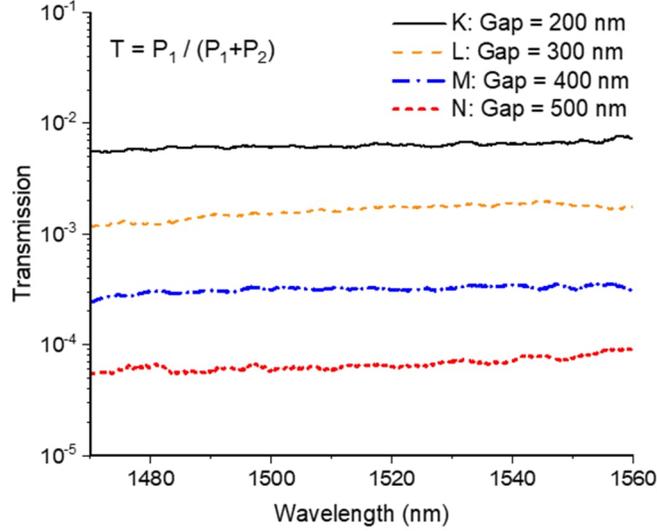

Fig. 2. Beam splitter transmission spectra for the gap widths used in resonators K, L, M and N. The transmission T is calculated as the ratio of the transmission output port power $P_1$ to the summed power of the transmission and reflection ports $P_1+P_2$ in Fig. 1(a).

Fig. 3 shows the measured spectra for the all-pass and add-drop versions of resonator K with a 200 nm splitter gap. The all-pass spectrum in Fig. 3(a) exhibits clearly defined resonances across the full laser tuning range. The free spectral range is $\Delta\lambda_{FSR} = 6.45$ nm, with resonance depths ranging from -8 to -13 dB. The through port spectrum for the add-drop version of resonator K in Fig. 3(b) shows a similar set of resonances with resonance depths from -6 dB to -9 dB. For both resonator structures the through port baseline insertion loss, including the parabolic collimators, is wavelength independent and approximately -1 dB. The variation in resonance depths arise because the externally reflected beam and the transmitted resonator output at mirror M1 can both be considered to be multimode beams consisting of many lateral modes of the Si slab waveguide. This is a basic operating difference from a waveguide directional coupler that evanescently couples light between two single mode waveguides. The coherent combination of the resonator output and reflected slab beams therefore may have a wavelength dependence when focused into the final single mode output waveguide. If the incident slab beam and internal resonator beam waist at mirror M1 were perfectly matched in shape, this effect would be mitigated because the input single mode waveguide profile would simply be reconstructed at the output waveguide. For the same reason, the background through port signals in Figs. 3(a) and (b) are wavelength independent, since the off-resonance through port signal is dominated by the reflected input beam from M1, which is simply re-focused back into the output waveguide. Fig. 3(c) shows the drop port spectrum for the same add-drop resonator as in Fig. 3(b). In this case the resonance peaks all have the same peak intensity since the drop port signal comprises only the resonator output coupling at mirror M2. Note that in the drop port spectrum in Fig. 3(c) also reveals a very weak second resonance comb at -30 dB lower peak power but almost the same FSR as the primary comb. This suggests that the planar equivalent of the asymmetric second order Gaussian cavity mode [16] is weakly excited, possibly due to small mismatch in the input and resonator mode overlaps at M1. The 3dB linewidth on the second comb resonances is approximately 70 pm, indicating this mode has a higher internal loss than the primary mode. Fig. 3(d) shows an expanded view of single resonance lines near λ=1521 nm. While they are nominally identical, the all-pass and add-drop resonator cavities are two different devices so the resonances do not occur at exactly the same wavelength.

Table 1. Measured and derived CGMR resonator properties[a]

| Resonator | K | L | M | N |
|---|---|---|---|---|
| Splitter Gap (nm) | 200 | 300 | 400 | 500 |
| Transmission | $6.6\times10^{-3}$ | $1.9\times10^{-3}$ | $3.2\times10^{-4}$ | $7.9\times10^{-5}$ |
| $\lambda_{res}$ (nm) | 1552.75 | 1552.79 | 1552.51 | 1545.13 |
| $\lambda_{3dB}$ (pm) | 24 | 14 | 8 | 5 |
| Q | 65000 | 111000 | 194000 | 310000 |
| $a^2$ | 0.989 | 0.990 | 0.993 | 0.995 |
| [b]Loss (dB/cm) > | -4.4 | -4.3 | -3.1 | -2.0 |
| [b]Reflectivity > | 0.9974 | 0.9975 | 0.9982 | 0.9988 |

[a]Based on analysis of measured resonator line shapes using Eq. 1.

[b]Minimum waveguide loss or mirror reflectivity compatible with the round-trip power attenuation $a^2$ given in the table.

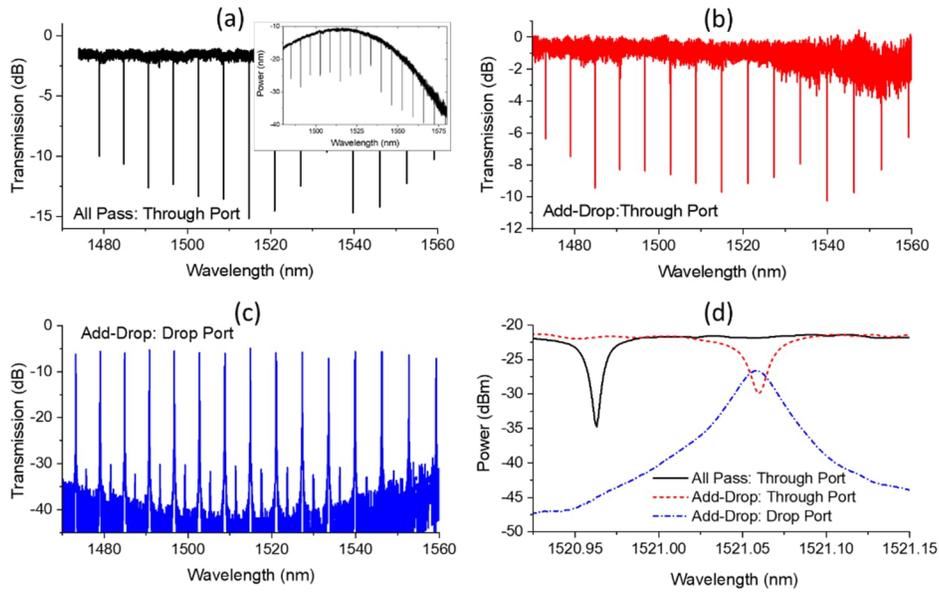

Fig. 3. The measured through and drop spectra for resonator K for (a) the all-pass, and in (b) and (c), the add-drop configuration. The spectra have been normalized to the reference waveguide and represent the resonator insertion loss. The inset of graph (a) shows the all-pass spectrum before normalization. (d) An expanded view of resonance lines near λ=1521 nm.

Fig. 4 shows an expanded view of selected drop port resonance lines near λ = 1550 nm for the add-drop resonators K, L, M, and N, as indicated in Table 1. The equivalent all-pass resonators have narrower FWHM resonance widths than the add-drop devices. However, the resonators are not critically coupled and have relatively shallow through port resonances compared with the signal background noise. Therefore, the line width analysis below is based on the drop port spectra of the add-drop devices for which the signal background is the detector noise floor. The drop-port resonance FWHM values given in Table 1 change from $\Delta\lambda_{3dB} = 24$ pm to $\Delta\lambda_{3dB} = 5$ pm with increasing beam splitter gap, and quality factors $Q = \lambda_{res}/\Delta\lambda_{3dB}$ values ranging from $Q = 65{,}000$ to $Q = 310{,}000$. The resonator Q at resonance wavelength $\lambda_{res}$ in a ring resonator can be expressed as

$$Q = \frac{\pi N_g L \sqrt{r_1 r_2\, a}}{\lambda_{res}(1 - r_1 r_2\, a)} \qquad (1)$$

where $N_g = 3.7$ is the Si slab waveguide group effective index as determined from the FSR [2], and $L = 100$ μm is cavity length. Here $r_1$ and $r_2$ are reflection coefficients of the beam splitters at mirror M1 and mirror M2, but they play the same role as the through coupling coefficients for the directional couplers in a conventional ring. The coefficient $a$ is the round-trip field propagation loss, so that $r_1 r_2 a$ represents the total round-trip field attenuation of the resonator. Using equation (1) and the measured resonance linewidths and beam splitter transmission, the internal properties of the CGBR given in Table 1 can be calculated. The round-trip power attenuation (excluding the outcoupling at the beam splitters), ranges from $a^2 = 0.989$ to $0.995$. Although it is not possible to separate cavity propagation loss from mirror reflection loss, one can estimate the worst-case limits on these two loss contributions. If mirror loss is assumed to be negligible, the highest resonator propagation losses range from -4.4 dB/cm for device K to -2.0 dB/cm for device N. This loss includes any absorption and defect scattering, loss arising from deviations of the circulating mode from an ideal Gaussian cavity, and any interaction of the Gaussian mode tails with the side walls of the slab waveguides. This result indicates that the cavity as designed supports a nearly perfect Gaussian mode, since a distance of 1 cm corresponds to approximately 100 round trips. On the other hand, if waveguide losses are assumed to be negligible, then the maximum loss per reflection is -0.012 dB. The lower limit on corner mirror power reflectivity is R > 99.7%, confirming that the mirrors are operating in TIR mode, and any imperfections at the mirror sidewalls do not create significant loss.

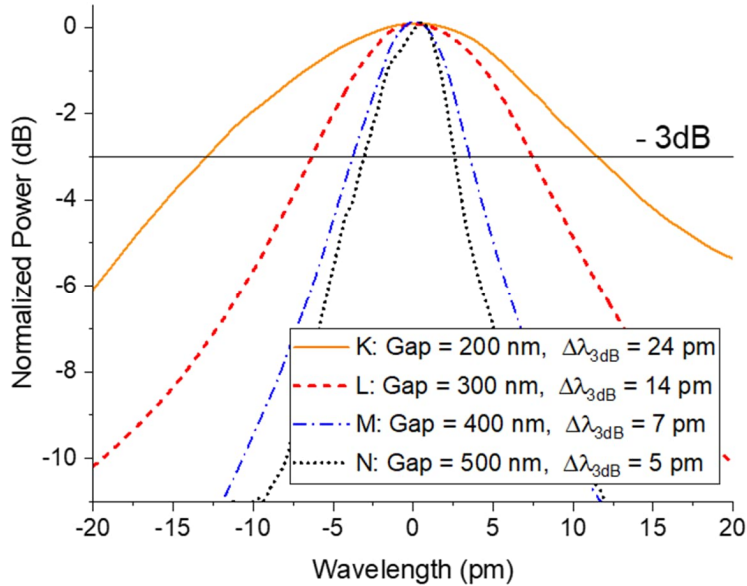

Fig. 4. Drop port resonance lines for resonators K, L, M, and N at the wavelengths given in Table 1. Each curve has been shifted in wavelength and power to facilitate comparison of line shapes. The horizontal -3 dB line intersects the spectra at the FWHM points.

## 4. Summary

In this work, a circulating Gaussian mode resonator (CGMR) design has been described and characterized. The CGMR resonator is implemented using 220 nm thick Si slab waveguide without lateral waveguide confinement. Instead, reflective structures are used to collimate and shape the propagating beam, and confine light in the form a Gaussian resonator cavity mode. Waveguide couplers are replaced by beam splitters that use evanescent coupling to split the incident light into a reflected and transmitted beam. Overall, the structure resembles a free space table top optical layout compressed into a 2-dimensional waveguide plane, as opposed to a traditional integrated optical circuit composed of very narrow single mode waveguides. Measured resonator linewidths of $\Delta\lambda_{3db}$=5 pm and Q factors up to Q=311000 have been measured.

There are several advantages to this design approach. By eliminating narrow (<500 nm) singe mode waveguides, waveguide losses due to sidewall roughness are reduced. Similarly, backscattering into counter-propagating modes by side wall defects is eliminated which should reduce the occurrence of resonator mode distortion and splitting. A qualitative survey of the measured spectra from resonators K, L, M, an N has found no sign of mode splitting. In contrast, in traditional ring resonators fabricated by a similar process tested in our laboratory, it is common that a significant fraction of the resonances will show mode splitting or distortion. The beam splitters used as input and output couplers in the CGMR device have a negligible wavelength dependence, so that a single CGMR device can provide uniform resonances over more than 100 nm wavelength range. In contrast, directional couplers have a usable coupling efficiency over a passband a few tens of nanometers wide with a peak coupling wavelength that is sensitive to fabrication variations. In the CGMR the circulating beam is several micrometers wide so local power is much lower than a similar single mode ring design. As a result, for a given input power, the local index perturbations due to self-heating and other nonlinear optical effects are much lower. This can be a particular advantage when resonators are used as sensors or as optical wavelength filters [15], where it is important to have precise resonance wavelength reproducibility that is independent of optical input power variations. For the 100 μW input power, no obvious self-heating resonance distortion was noted in any of the CGMR devices. At a higher power of 1 mW, device K did exhibit a self-heating resonance shift on the order of 3 pm. A 100 μm long Si ring resonator reported in Ref. [15] exhibited a self-heating shift of almost 20 pm at the same input power, but the caveat is that comparisons of such different devices can only be qualitative. Lastly, the slab waveguide CGMR resonator presents wide flat (100) Si upper surface to the ambient environment that should be more suitable for well controlled surface functionalization or other types of surface preparation for sensor and other applications [8]. This is in contrast to single mode waveguides where much of the Si surface area is on the vertical sidewalls which have no specific crystal orientation and have residual near-surface etch damage and roughness.

The CGMR resonator performance does depend on the mode matching of the incident slab beam shape and the resonator beam waist at the input beam splitter. This will determine overall resonator transmission near resonance, and in particular the detailed characteristics of the through-port spectrum. In this work, simple parabolic beam collimator and a circular concave corner mirror set at 45° incidence were used to build the resonator. This configuration already gives off-resonance through port insertion losses better than -3 dB, and comparable Q-factors to the best Si ring resonators measured in our laboratory. With further design optimization for planar slab mode beam shaping and collimation, we anticipate that resonators with improved output characteristics, a wide range of cavity lengths, and high Q values can be achieved.

**Disclosures.** The authors declare no conflicts of interest.

**Data availability.** Data underlying the results presented in this paper are not publicly available at this time but may be obtained from the authors upon reasonable request.

## References

1. A. Yariv, "Universal relations for coupling of optical power between microresonators and dielectric waveguides" Electronics Lett. **36**(4), 321-322 (2000).
2. W. Bogaerts, P. De Heyn, T. Van Vaerenbergh, K. De Vos, S. Kumar Selvaraja, T. Claes, P. Dumon, P. Bienstman, V. Van Thourhout, and R. Baets. "Silicon microring resonators," Laser and Phot. Rev. **6**, 47-73 (2011).
3. D.O.M. de Aguiar, M. Milanizadeh, E. Guglielmi, F. Zanetto, G. Ferrari, M. Sampietro, F. Morichetti, and A. Melloni. "Automatic tuning of silicon photonics microring filter array for hitless reconfigurable add–drop." J. Lightwave Technol. **37**(16), 3939-3947 (2019).
4. Hsu, Wei-Che, Nabila Nujhat, Benjamin Kupp, John F. Conley Jr, and Alan X. Wang. "On-chip wavelength division multiplexing filters using extremely efficient gate-driven silicon microring resonator array." Scientific Reports **13** (1), 5269 (2023).
5. Q. Xu, B. Schmidt, J. Shakya, and M. Lipson. "Cascaded silicon micro-ring modulators for WDM optical interconnection," Opt..Express **14**(20), 9431-9436 (2006).
6. Rosenberg, J. C., W. M. J. Green, S. Assefa, D. M. Gill, T. Barwicz, M. Yang, S. M. Shank, and Y. A. Vlasov. "A 25 Gbps silicon microring modulator based on an interleaved junction." Opt. Express **20** (24) 26411-26423 (2012).
7. S. Dedyulin, Z. Ahmed, G. Machin, "Emerging technologies in the field of thermometry," Meas. Sci. Technol. **33**, 092001, 1-26 (2022).
8. Janz, S., D-X. Xu, M. Vachon, N. Sabourin, P. Cheben, H. McIntosh, H. Ding et al. "Photonic wire biosensor microarray chip and instrumentation with application to serotyping of Escherichia coli isolates." Optics express **21** (4), 4623-4637 (2013).
9. N. Klimov, T.P. Purdy, Z. Ahmed, "Towards replacing resistance thermometry with photonic thermometry," Sens. Actuators A Phys. **269**, 308–312 (2018).
10. S. Dedyulin, A. Todd, S. Janz, D.-X. Xu, S. Wang, M. Vachon and J. Weber, "Packaging and precision testing of fiber-Bragg-grating and silicon ring-resonator thermometers: current status and challenges," Meas. Sci. Technol. **31**, 074002, 1-7 (2020).
11. R. Eisermann, S. Krenek, G. Winzer, and S. Rudtsch, "Photonic contact thermometry using silicon ring resonators and tuneable laser-based spectroscopy," Technisches Messen **88**, 640–654 (2021).
12. A. Li, T. Van Vaerenbergh, P. De Heyn, P. Bienstman, and W. Bogaerts, "Backscattering in silicon microring resonators: a quantitative analysis," Laser Photonics Rev. **10** (3), 420–431 (2016).
13. B.E. Little, J.-P. Laine, "Surface Roughness induced contradirectional coupling in ring and disk resonators," Opt. Lett. **22**(1), 4-6 (1997).
14. D.-X. Xu, A. Densmore, P. Waldron, J. Lapointe, E. Post, A. Delâge, S. Janz, P. Cheben, J.H. Schmid, and B. Lamontagne, "High bandwidth SOI photonic wire ring resonators using MMI couplers," Opt. Express **15**, 3149-3155 (2007)
15. S. Janz, S. Dedyulin, D.-X. Xu, M. Vachon, S. Wang, R. Cheriton, and J. Weber, "Measurement accuracy in silicon photonic ring resonator thermometers: identifying and mitigating intrinsic impairments," Optics Express **32**, (1), 551-575 (2023).
16. A. Yariv, *Quantum Electronics*, Chap. 6, 2nd edition, John Wiley and Sons, New York, (1975).
17. H. Xu, Y. Qin, G. Hu, and H. K. Tsang, "Compact integrated mode-size converter using a broadband ultralow-loss parabolic-mirror collimator," Optics Lett. **48**(2), 327-330 (2023).
18. L. Moreno-Pozas, M. Barona-Ruiz, R. Halir, J. de-Oliva-Rubio, J. Rivas-Fernández, I. Molina-Fernández, J. G. Wangüemert-Pérez, and A. Ortega-Moñux, "Parabolic dielectric reflector for extreme on-chip spot-size conversion with broad bandwidth," Optics Lett. **50**(4), 1073-1076 (2025).
19. Y. Wang, X. Wang, J. Flueckiger, H. Yun, W. Shi, R. Bojko, N. A. F. Jaeger, and L. Chrostowski, "Focusing sub-wavelength grating couplers with low back reflections for rapid prototyping of silicon photonic circuits," Optics Ex. **22**(17), 20652 (2014).
20. Applied Nanotools Inc., *"NanoSOI Silicon PDK Component Documentation,"* 2023.